\documentclass[prl,showpacs,amsmath,amssymb, twocolumn, 10pt]{revtex4}
\usepackage{enumerate}
\usepackage{color}
\usepackage{amsthm}
\usepackage{dcolumn}
\usepackage{bm}
\usepackage{graphicx}

\usepackage{soul, color}

\begin{document}
\newtheorem{corollary}{Corollary}
\newtheorem{conjecture}{Conjecture}
\newtheorem{definition}{Definition}
\newtheorem{example}{Example}
\newtheorem{lemma}{Lemma}
\newtheorem{proposition}{Proposition}
\newtheorem{theorem}{Theorem}
\newtheorem{fact}{Fact}
\newtheorem{property}{Property}

\newcommand{\nc}{\newcommand}
\nc{\bra}[1]{\langle#1|} \nc{\ket}[1]{|#1\rangle} \nc{\proj}[1]{|
#1\rangle\!\langle #1 |} \nc{\ketbra}[2]{|#1\rangle\!\langle#2|}
\nc{\braket}[2]{\langle#1|#2\rangle} \nc{\norm}[1]{\lVert#1\rVert}
\nc{\abs}[1]{|#1|} \nc{\lar}{\leftarrow} \nc{\rar}{\rightarrow}
\nc{\ox}{\otimes} \nc{\op}[2]{|#1\rangle\!\langle#2|}
\nc{\ip}[2]{\langle#1|#2\rangle} \nc{\dg}{\dagger}
\nc{\fract}{\theta_l}
\nc{\fracts}{\overset{N}{\underset{l=1}{\Sigma}}\theta_l}
\nc{\fractml}{\left((-1)^{m_l}\theta_l\right)}
\nc{\fractmls}{(\overset{N}{\underset{l=1}{\Sigma}}(-1)^{m_l}\theta_l)}
\nc{\fractmlst}{[(\overset{N}{\underset{l=1}{\Sigma}}(-1)^{m_l}\theta_l)/2]}


\title{Perfect NOT  and  conjugate transformations}

\author{Fengli Yan$^1$}
 \email{flyan@hebtu.edu.cn}
 \author{Ting Gao$^2$}
 \email{gaoting@hebtu.edu.cn}
 \author{Zhichao Yan$^3$}
 \affiliation{$^1$ College of Physics
Science and Information Engineering and Hebei Advanced Thin Films
Laboratory,
Hebei Normal University, Shijiazhuang 050024, China\\
$^2$ College of Mathematics and Information Science,
Hebei Normal University, Shijiazhuang 050024, China\\
$^3$ College of Software, Hebei Normal University, Shijiazhuang
050024, China }
\date{\today}

\begin{abstract}
{This paper reports on a study of the perfect NOT, probabilistic perfect NOT and conjugate transformations. The perfect NOT transformation criteria, two necessary and sufficient conditions for realizing a perfect NOT transformation on a quantum state set $S$ of a qubit, are  obtained. Furthermore, this paper discusses a probabilistic perfect NOT transformation (gate)
when there is no perfect NOT transformation on a state set $S$
 and the construction of a probabilistic perfect NOT machine (gate) by a general unitary-reduction operation is shown. With a postselection
 of measurement outcomes, the probabilistic NOT gate yields perfect orthogonal complements of the input states.
  We also generalize the perfect NOT transformation to
the conjugate transformation  in the multi-level quantum system and a lower bound of the best possible efficiencies attained by a probabilistic perfect conjugate transformation are obtained. }
\end{abstract}
 \maketitle
\section{1. introduction}

The basic building block of any classical information processor is the single  bit,  which is prepared in one of two possible states, denoted  0 or 1. However, quantum information consists of qubits, each of which has the luxury of being in a superposition of the  0 and 1 states. Since there are an infinite number of superposition states, quantum systems have a much richer and more interesting existence than their classical counterparts. The superposition of states also makes the properties of quantum information quite different from that of its classical counterpart. Whereas the copying of classical information presents no difficulties, owing to the linearity of quantum mechanics, there is a quantum no-cloning theorem \cite {Wootters-Zurek,Dieks} which asserts that it is impossible to construct a device that will perfectly copy an
arbitrary (unknown) state of a two-level particle. However, the quantum no-cloning theorem does not rule out the possibility of either imperfect cloning \cite {Buzek-Hillery, Scarani-Iblisdir-Gisin-Acin} or probabilistic cloning \cite {Duan-Guo}. Some applications of  cloning have been presented \cite {Galvo-Hardy, Gao-Yan-Wang, Gao-Yan-Wang-Li}. With the progress of a quantum information theory, quantum cloning has become a quite interesting field.

There is another difference between classical and quantum
information systems. It is very easy to complement a classical bit, i.e., to
change the value of a bit, a 0 to a 1 and vice versa. Usually this
operation can be accomplished by a NOT transformation (gate).
However, in quantum information systems, changing an unknown state
$|\Psi\rangle=\alpha|0\rangle+\beta|1\rangle$ of a qubit to its
orthogonal complement $|\Psi^\bot\rangle=\alpha^*|1\rangle-\beta^*|0\rangle$
that is orthogonal to $|\Psi\rangle$ (i.e. inverting the state of a
two-level quantum system) is impossible \cite{Bechmann-Gisin,Buzek-Hillery-Werner}.  The result is that one can not design a device that will take an arbitrary qubit and transform it into its orthogonal qubit. This is because complex conjugation of the coefficients in the NOT
transformation of a qubit must be accomplished by an
antiunitary transformation and cannot be performed by a unitary
one. In other words, it is impossible to achieve the perfect NOT
gate in quantum information systems.

However, the NOT  transformation can be achieved on some states while leaving other states unchanged. Alternatively, there can be a transformation operation that approximates, at best, the NOT gate on all states, called the universal NOT gate
\cite{Bechmann-Gisin,Buzek-Hillery-Werner}.  In fact, the output of a quantum cloning
machine, the ancilla, carries the optimal anticlone of the input state so the universal NOT gate can be accomplished as a by-product of cloning \cite { Scarani-Iblisdir-Gisin-Acin}.

A combination of unitary evolution together with measurements is
an important method in  quantum information processing and often
achieves very interesting results. It has been used in quantum programming \cite{Nielsen-Chuang}, the purification of entanglement \cite{Bennett-Brassard-Popescu-Schumacher-Smolin-Wootters},  quantum
teleportation \cite{Bennett-Brassard-Crepeau-Jozsa-Peres-Wootters} and the preparation of quantum states \cite{Brune-Haroche-Raimond-Davidovich-Zagury}.  Recently, by using
this method, Duan and Guo designed a probabilistic quantum cloning
machine \cite{Duan-Guo}. With a postselection of the measurement
results, the machine outputs perfect copies of the input states.

In this paper, the perfect NOT, probabilistic perfect NOT and conjugate transformations are investigated.
We present the criteria for a perfect NOT transformation on a quantum state set $S$ of a qubit. Two necessary and sufficient conditions for realizing a perfect NOT transformation on $S$ are derived and this paper discusses how to  build a device to achieve probabilistic perfect NOT transformations when there is no perfect NOT transformation on the state set  $S$. With certain nonzero probabilities of success, this device  transforms an arbitrary unknown input state   into its  orthogonal complement. We also generalize the  probabilistic NOT transformation to the conjugate transformation in the multi-level quantum system. Furthermore, the lower bound of the best possible efficiencies attained by a probabilistic perfect conjugate transformation is obtained.

\section{2. Conditions required for perfect NOT transformations}

In order to aid the analysis, we first state a Lemma of
Duan and Guo \cite {Duan-Guo}:

\emph{Lemma.}  If two sets of states $|\phi_1\rangle,
|\phi_2\rangle, \cdots, |\phi_n\rangle$ and $|\tilde{\phi_1}\rangle,
|\tilde{\phi_2}\rangle, \cdots, |\tilde{\phi_n}\rangle$ satisfy the
condition
\begin{equation}\begin{array}{cc}
\langle\phi_i|\phi_j\rangle=\langle\tilde{\phi_i}|\tilde{\phi_j}\rangle,& (i,j=1,2,\cdots, n),
\end{array}\end{equation}
then there exists a unitary operator $U$ such that $U|\phi_i\rangle=|\tilde{\phi}_i\rangle,$ $(i=1,2,\cdots,n)$.

Let $S=\{|\Psi_1\rangle, |\Psi_2\rangle, \cdots, |\Psi_n\rangle\}$ be a set of states of a qubit. 
When the quantum states of $S$ satisfy
\begin{equation}\begin{array}{cc} \langle \Psi_i|\Psi_j\rangle=\langle \Psi_i^\perp| \Psi_j^\perp\rangle,&
 (i,j=1,2,\cdots,n),\end{array}
\end{equation}
based on Lemma  we can find a unitary transformation $U$ such that
$\begin{array}{ccc} |\Psi^\perp_i\rangle=U|\Psi_i\rangle & {\rm for}
& i=1,2,\cdots, n.\end{array}$

It is easy to see that $\langle
\Psi_i^\perp|\Psi_j^\perp\rangle=(\alpha_i\langle 1| -\beta_i\langle
0|) (\alpha_j^*|1\rangle-\beta_j^*|0\rangle)=
\alpha_i\alpha_j^*+\beta_i\beta_j^*=\langle \Psi_i|\Psi_j\rangle^*$, which shows  that
the condition Eq.(2) is equivalent to $\langle
\Psi_i|\Psi_j\rangle=\langle \Psi_i|\Psi_j\rangle^*$ for
$i,j=1,2,\cdots, n$.  This, in turn, implies that all inner-products of the quantum
states in the set $S$ are real. Hence we arrive at the following
conclusion:

\emph{Theorem 1.} Suppose that $S=\{|\Psi_1\rangle, |\Psi_2\rangle, \cdots,
|\Psi_n\rangle\}$ is the set of quantum states. Then a perfect NOT transformation (gate) $U$ on the set $S$ can be realized by a unitary transformation (i.e., there is a unitary transformation $U$ such that $U|\Psi_i\rangle=|\Psi_i^\perp\rangle$) if and only if 
\begin{equation}\label{}
\langle \Psi_i|\Psi_j\rangle=\langle \Psi_i|\Psi_j\rangle^*, ~~ i,j=1,2,\cdots, n.
\end{equation}

It turns out that if $S$ contains all points of a Bloch sphere \cite{Nielsen-Chuang-book}  of a qubit, one can not realize the perfect NOT transformation on the set $S$.

 Obviously, in Theorem 1 we only consider  the case without an ancilla (probe). Now, let us
 introduce a probe $P$ with the initial state $|P^{(0)}\rangle$.  By Lemma, if
\begin{eqnarray}\label{GaoYan2}
 & &\langle P^{(0)}|\langle\Psi_i|\Psi_j\rangle|P^{(0)}\rangle\nonumber\\
=&&\langle\Psi_i|\Psi_j\rangle\nonumber\\
=&&\langle\Psi_i^\perp|\Psi_j^\perp\rangle\langle
P^{(i)}|P^{(j)}\rangle\nonumber\\
=&&\langle\Psi_i|\Psi_j\rangle^*\langle P^{(i)}|P^{(j)}\rangle,
\end{eqnarray}
for arbitrary $i,j=1,2,\cdots, n$,  then there exists a unitary
transformation $U$, such that
\begin{equation}\label{important}
U(|\Psi_i\rangle|P^{(0)}\rangle)=|\Psi_i^\perp\rangle|P^{(i)}\rangle.
\end{equation}
It means we can realize the perfect NOT transformation on the
quantum state set  $S=\{|\Psi_1\rangle, |\Psi_2\rangle, \cdots,
|\Psi_n\rangle\}$ with the assistance of the ancilla (probe).

Next, we discuss the case in which $\langle \Psi_i|\Psi_j\rangle\neq
0$ for arbitrary $i,j=1,2,\cdots,n$. Let
$\langle
\Psi_i|\Psi_j\rangle=t_{ij}\texttt{e}^{\texttt{i}\theta_{ij}}$, and
$\langle
P^{(i)}|P^{(j)}\rangle=p_{ij}\texttt{e}^{\texttt{i}\varphi_{ij}}$. Here $0<t_{ij},p_{ij}\leq1$, and $0\leq \theta_{ij},\varphi_{ij}<2\pi$.
 Eq.(\ref{GaoYan2})  is then equivalent to
\begin{equation}
t_{ij}\texttt{e}^{\texttt{i}\theta_{ij}}=t_{ij}\texttt{e}^{-\texttt{i}\theta_{ij}}p_{ij}\texttt{e}^{\texttt{i}\varphi_{ij}}.
\end{equation}
It implies that
\begin{equation}\label{GaoYan3}
\begin{array}{cc}p_{ij}=1,& 2\theta_{ij}=\varphi_{ij}+2k_{ij}\pi,
\end{array}\end{equation}
where $k_{ij}=0$ or 1.
Because $p_{1j}=1$, there must be
\begin{equation}
|P^{(i)}\rangle=\texttt{e}^{\texttt{i}\varphi_i}|P^{(1)}\rangle
\end{equation}
for $i=2,3,\cdots,n$, and $\varphi_1=0$. It follows that
\begin{equation}
\varphi_{ij}=\varphi_j-\varphi_i+2m_{ij}\pi,
\end{equation}
where $m_{ij}=0$ or 1.
By Eq.(\ref{GaoYan3}), we obtain
\begin{equation}
\varphi_j-\varphi_i=2\theta_{ij}-2(k_{ij}+m_{ij})\pi.
\end{equation}
Thus,
\begin{equation}
\varphi_j=2\theta_{1j}-2(k_{1j}+m_{1j})\pi.
\end{equation}
Therefore,
\begin{equation}\label{gaoyan4}
\begin{array}{c}
  \theta_{ij}=\theta_{1j}-\theta_{1i}+(k_{ij}+m_{ij}-k_{1j}-m_{1j}+k_{1i}+m_{1i})\pi,\\
  i,j=1,2,\cdots,n, 
\end{array}
\end{equation}
 which implies that
\begin{equation}\label{yangao1}
\begin{array}{c}
  \theta_{lj}-\theta_{li}=\theta_{ij}+(k_{lj}+m_{lj}-k_{li}-m_{li}-k_{ij}-m_{ij})\pi, \\
  i,j,l=1,2,\cdots,n;~k_{ij},m_{ij}=0 ~ \text{or} ~ 1.
\end{array}
\end{equation}

Furthermore, starting from  Eq.(\ref{yangao1}) we can reverse the process. This means that if the quantum state set $S=\{|\Psi_1\rangle, |\Psi_2\rangle, \cdots, |\Psi_n\rangle\}$ satisfies $\langle\Psi_i|\Psi_j\rangle\neq 0$ and Eq.(\ref{yangao1}), one can find a unitary transformation $U$ such that Eq.(\ref{important}) hold and a perfect NOT transformation can be realized.

Based on the above argument we obtain following conclusion:

\emph{Theorem 2.} Suppose that the quantum state set $S=\{|\Psi_1\rangle, |\Psi_2\rangle, \cdots, |\Psi_n\rangle\}$ satisfies $\langle\Psi_i|\Psi_j\rangle\neq 0$. Then, a perfect NOT transformation
(gate) on the state set $S$ can be realized by a unitary transformation acting on the system and a probe if and only if Eq.(\ref{yangao1}) hold.

Note that, when the quantum state set $S$ contains only two quantum states $|\Psi_1\rangle,|\Psi_2\rangle$, Eq.(\ref{yangao1}) can always hold. Therefore, the perfect NOT transformation (gate) on
the state set $S$ of two arbitrary quantum states $|\Psi_1\rangle,|\Psi_2\rangle$ can always be realized.

Clearly, if there are no quantum states $|P^{(1)}\rangle,|P^{(2)}\rangle, \cdots, |P^{(n)}\rangle$ satisfying Eq.(4) for the  quantum state set $S$, then one can not design a perfect NOT gate for this state set $S$.
In this case one can only consider the universal-NOT or the probabilistic perfect NOT gate. As the universal-NOT gate has been well studied \cite {Bechmann-Gisin,Buzek-Hillery-Werner}, in the next section, we will only discuss the probabilistic perfect NOT gate in detail.

\section{3. Probabilistic perfect NOT transformation}

The definition of a probabilistic perfect NOT gate is that for a quantum state set
$S=\{|\Psi_1\rangle, |\Psi_2\rangle, \cdots, |\Psi_n\rangle\},$ there
is a unitary transformation together with a measurement, which when combined with a
postselection of measurement results, makes an arbitrarily unknown input quantum state $|\Psi_i\rangle$ transform into its orthogonal complement $|\Psi_i^\perp\rangle$ with certain nonzero probability of success.
That is, for a quantum state set $S=\{|\Psi_1\rangle, |\Psi_2\rangle, \cdots,|\Psi_n\rangle\}$,  if there exists a unitary operation $U$ and a measurement $M$, which together yield the following evolution:
\begin{equation}\label{yan1}
|\Psi_i\rangle\begin{array}{c} U+M\\\longrightarrow
\end{array}
|\Psi_i^\bot\rangle,\end{equation}
then a probabilistic NOT gate is said to have been built. The combination of a unitary evolution operation and a measurement is very general and can be used to describe any operation in quantum mechanics \cite {Kraus}.

Obviously, we can not build a probabilistic NOT gate for any arbitrary quantum state set $S=$ $\{|\Psi_1\rangle, |\Psi_2\rangle, \cdots, |\Psi_n\rangle\}$, so it is very important to find the conditions
that the quantum state set $S$ should be satisfied in order to construct a probabilistic perfect NOT gate.

The unitary evolution of the qubit  $A$ and probe $P$ can be described by the following equation
\begin{equation}\label{yan}
\begin{array}{c}
U(|\Psi_i\rangle|P_0\rangle)=\sqrt
{\gamma_i}|\Psi_i^\perp\rangle|P^{(i)}\rangle+\sqrt
{1-\gamma_i}|\Phi^{(i)}_{AP}\rangle,\\
(i=1,2,...,n),
\end{array}
\end{equation}
where $|P_0\rangle$ and $|P^{(i)}\rangle$  are normalized states of
the probe $P$ (not generally orthogonal) and
$|\Phi^{(1)}_{AP}\rangle$, $|\Phi^{(2)}_{AP}\rangle$, $\cdots$, and
$|\Phi^{(n)}_{AP}\rangle$ are $n$ normalized states of the composite
system $AP$ (not generally orthogonal). We assume that in
Eq.(\ref{yan}) the coefficients before the states
$|\Psi_i^\perp\rangle|P^{(i)}\rangle$,  and
$|\Phi^{(i)}_{AP}\rangle$ are  positive real numbers.  Let $S_0$ be
the subspace spanned by the states $|P^{(1)}\rangle$,
$|P^{(2)}\rangle$, $\cdots$, $|P^{(n)}\rangle$. In order to realize
the probabilistic perfect NOT transformation, we must require that
after the unitary evolution a measurement of the probe with a
postselection of the measurement results should project its state
into the subspace $S_0$. After this projection, the
state of the system $A$ should be $|\Psi_i^\perp\rangle$. Therefore,
all of the states $|\Phi^{(i)}_{AP}\rangle$,  lie in a space orthogonal to $S_0$ and can be represented  by the following
equation
\begin{equation}\begin{array}{c}
|P^{(i)}\rangle\langle P^{(i)}|\Phi_{AP}^{(j)}\rangle=0,\\
(i,j=1,2,...,n). \end{array}\end{equation}
 With  above restriction,
inter-inner-products of Eq.(\ref{yan}) yield the following matrix
equation
\begin{equation}\label{gaoyan}
X^{(1)}=\sqrt \Gamma X^{(\perp)}_P\sqrt {\Gamma^+}+\sqrt
{E_n-\Gamma}Y\sqrt {E_n-\Gamma^+},
\end{equation}
where $X^{(1)}=[\langle \Psi_i|\Psi_j\rangle]$,
$Y=[\langle\Phi^{(i)}_{AP}|\Phi^{(j)}_{AP}\rangle]$,
$X^{(\perp)}_P=[\langle \Psi_i^\perp|\Psi_j^\perp\rangle\langle
P^{(i)}|P^{(j)}\rangle]$ are $n\times n$ matrices and $E_n$ is the $n\times n$ identity matrix.
The diagonal efficiency matrix $\Gamma$ is
defined by $\Gamma={\rm diag}(\gamma_1, \gamma_2,...,\gamma_n)$;
therefore, ${\sqrt \Gamma}={\sqrt {\Gamma^+}}= {\rm diag}({\sqrt
\gamma_1}, {\sqrt \gamma_2},...,{\sqrt \gamma_n})$.
According to result of Duan and Guo \cite {Duan-Guo},  $Y$ is a
positive-semidefinite matrix.  Thus,  $\sqrt {E_n-\Gamma}Y\sqrt
{E_n-\Gamma^+}$ is  also a positive-semidefinite matrix. Based on
Eq.(\ref{gaoyan}), $X^{(1)}-\sqrt \Gamma X^{(\perp)}_P\sqrt
{\Gamma^+}$ is a positive-semidefinite matrix. Conversely, if
$X^{(1)}-\sqrt \Gamma X^{(\perp)}_P\sqrt {\Gamma^+}$ is a
positive-semidefinite matrix, one can choose
$|\Phi^{(i)}_{AP}\rangle$ such that Eq.(\ref{gaoyan}) holds. By
Lemma  the states $|\Psi_1\rangle, |\Psi_2\rangle, . . . ,$ and
$|\Psi_n\rangle$   are able to be probabilistically transformed to
their respective orthogonal complement states. Thus we have the following theorem:

\emph{Theorem 3}. The states $|\Psi_1\rangle, |\Psi_2\rangle, . . .,$ and $|\Psi_n\rangle$
  can be probabilistically perfectly transformed to their respective orthogonal  complement states if and only if there exist  a diagonal
  positive-definite matrix $\Gamma$ and $|P^{(i)}\rangle$ ($i=1,2,...,n$) such that   the matrix $X^{(1)}-\sqrt
 \Gamma X^{(\perp)}_P\sqrt {\Gamma^+}$ is positive-semidefinite. Here 
$X^{(1)}=[\langle \Psi_i|\Psi_j\rangle]$ and 
$X^{(\perp)}_P=[\langle \Psi_i^\perp|\Psi_j^\perp\rangle\langle
P^{(i)}|P^{(j)}\rangle]$ are $n\times n$ matrices, and $|P^{(i)}\rangle$ $(i=1,2,\cdots,n)$ are quantum states of a probe.

Theorem 3 is  very general, and for the linearly independent quantum
 state set $S$ we have the conclusion:

\emph{Theorem 4.} The states secretly chosen from the set $S=\{
|\Psi_1\rangle, |\Psi_2\rangle, \cdots, |\Psi_n\rangle\}$ can be
probabilistically transformed into their respective orthogonal complements by a general
unitary-reduction operation, if  $ |\Psi_1\rangle, |\Psi_2\rangle,
\cdots,$ and $|\Psi_n\rangle$ are linearly independent.

\emph{Proof}: Suppose the Hilbert space of the probe $P$ is an
$n_p$-dimensional  space, where $n_p\geq n+1$. We use
$|P_0\rangle,|P_1\rangle, . . . ,$ and $|P_n\rangle$ to denote $n+1$
orthonormal states of a probe $P$. If there exists a unitary
operator $U$ that satisfies
\begin{equation}\label{zhichao}\begin{array}{c}
U(|\Psi_i\rangle|P_0\rangle)=\sqrt \gamma_i |\Psi_i^\perp\rangle
\texttt{e}^{\texttt{i}\varphi_i}|P_0\rangle +\sum_{j=1}^n c_{ij}|\Phi^{(j)}\rangle |P_j\rangle, \\
(i=1,2,\cdots, n),
\end{array}\end{equation}
where $|\Phi^{(j)}\rangle$ $(j=1,2,\cdots,n)$ stand for $n$ normalized states of the
system (not generally orthogonal) and $\varphi_i$ are real
numbers, then after the evolution a measurement of the probe $P$ is
followed.    Eq.(\ref{zhichao}) is a special case of Eq.(\ref{yan}). The NOT transformation is successful, and the output
state of the system is $|\Psi_i^\perp\rangle$, if and only if the
measurement outcome of the probe is $|P_0\rangle$. Evidently, the
probability of success ( obtaining $|P_0\rangle$) is $\gamma_i$. For
any input state $|\Psi_i\rangle$, the probabilistic NOT device
should succeed  with a nonzero probability. This, in turn, implies that all of
the $\gamma_i$ must be positive real numbers. Hence, the evolution
(\ref{yan1}) can be realized  if Eq.(\ref{zhichao}) holds with
positive efficiencies $\gamma_i$. The $n\times n$
inter-inner-products of Eq.(\ref{zhichao}) yield the equation
\begin{equation}\label{zhaochao2} X^{(1)}={\sqrt \Gamma} X^{(\perp)}{\sqrt
{\Gamma^+}}+CC^+,
\end{equation}
where the $n\times n$ matrices $C=[c_{ij}]$,  $X^{(1)}=[\langle
\Psi_i|\Psi_j\rangle],$ and
$X^{(\perp)}=[\texttt{e}^{\texttt{i}(\varphi_j-\varphi_i)}\langle
\Psi_i^\perp|\Psi_j^\perp\rangle]=[\texttt{e}^{\texttt{i}(\varphi_j-\varphi_i)}\langle
\Psi_i|\Psi_j\rangle^*]$.  By considering
Lemma we know that if there exists a diagonal positive-definite
matrix $\Gamma$
 satisfied Eq.(\ref{zhaochao2}), then one can realize  the unitary evolution (\ref{yan1}).

Duan and Guo \cite{Duan-Guo} have shown that: If $n$ states
$|\Psi_1\rangle, |\Psi_2\rangle, . . . ,$ and $|\Psi_n\rangle$ are
linearly independent, the matrix
$X^{(1)}=[\langle\Psi_i|\Psi_j\rangle]$ is positive definite.

Suppose  that  the minimum eigenvalue of $X^{(1)}$ is $c$ and the maximum
eginvalue of $X^{(\perp)}$ is $d$. Then there must exist a
positive number $\varepsilon$ such that
\begin{equation}
c-\varepsilon d> 0.
\end{equation}
Let $B=(b_1, b_2, ..., b_n)^T$ be an arbitrary nonzero $n$ dimensional vector. Then
\begin{equation}
B^+(c-\varepsilon d)B> 0.
\end{equation}
It also follows that
\begin{equation}
B^+(cE-\varepsilon dE)B> 0,
\end{equation}
where  $E$ is the $n\times n$ identity matrix.
Presume  that $X^{(1)}$ and $X^{(\perp)}$ are diagonalized  by the unitary matricies $U$ and $V$, respectively. Eq.(22) can then be rewritten as
\begin{equation}
B^+(U^+cU-\varepsilon V^+dV)B> 0.
\end{equation}
We use $c_1, c_2, ..., c_n$ and $d_1,d_2,...,d_n$ to denote the
eginvalues of matrixes $X^{(1)}$ and $X^{(\perp)}$, respectively. It
is easy to deduce
\begin{equation}
B^+ [U^+{\rm diag} (c_1, ..., c_n)U-\varepsilon V^+{\rm diag}
(d_1,...,d_n)V]B> 0.
\end{equation}
That is,
\begin{equation}
B^+[X^{(1)}-\varepsilon X^{(\perp)}]B> 0.
\end{equation}

Obviously, there must be a diagonal matrix $\Gamma={\rm
diag}(\gamma_1, \gamma_2,...,\gamma_n)$ with $\gamma_i>0$ that
satisfies
\begin{equation}
\varepsilon X^{(\perp)}=\sqrt \Gamma X^{(\perp)}\sqrt {\Gamma^+}.
\end{equation}
Therefore, there is a diagonal matrix $\sqrt \Gamma$ such that
\begin{equation}
 X^{(1)}-\sqrt \Gamma X^{(\perp)}\sqrt {\Gamma^+}
\end{equation}
is positive definite.

Suppose  that the unitary matrix $W$ diagonalizes the Hermitian
matrix $X^{(1)}-{\sqrt \Gamma} X^{(\perp)}{\sqrt {\Gamma^+}}$ , that
is,
\begin{equation}
W(X^{(1)}-{\sqrt \Gamma} X^{(\perp)}{\sqrt {\Gamma^+}})W^+={\rm
diag}(m_1,m_2,...,m_n),\end{equation} where all of the eigenvalues
$m_1, m_2, . . . ,  m_n$ are positive real numbers. We can then
choose the matrix $C$ in Eq.(\ref{zhaochao2}) to be
\begin{equation}
C =W^+ {\rm diag}(\sqrt {m_1}, \sqrt {m_2}, . . . ,\sqrt
{m_n})W.\end{equation} Thus, there exists  a
diagonal positive definite efficiency matrix $\Gamma$ such that Eq.(19) holds and the proof of Theorem 4 is complete.

Next we consider probabilistic perfect NOT transformation of the
quantum state set $\{|\Psi_1\rangle, |\Psi_2\rangle,
|\Psi_3\rangle\}$. Suppose that $|\Psi_1\rangle$ and
$|\Psi_2\rangle$ are linearly independent and that
\begin{equation}
|\Psi_3\rangle=\alpha|\Psi_1\rangle+\beta|\Psi_2\rangle.
\end{equation}
Here  $\alpha$ and $\beta$  satisfy the normalizing condition
\begin{equation}\label{gaoyan55}
\alpha\alpha^*+\beta\beta^*+\alpha^*\beta\langle\Psi_1|\Psi_2\rangle+\alpha\beta^*\langle\Psi_2|\Psi_1\rangle=1.
\end{equation}

 From  Theorem 4, there exists a unitary
transformation $U$ such that
\begin{eqnarray}\label{gaoyan5}
&& U(|\Psi_1\rangle|P_0\rangle)=\sqrt
{\gamma_1}|\Psi_1^\perp\rangle|P_0\rangle+\sqrt
{1-\gamma_1}|\Phi^{(1)}_{AP}\rangle,\nonumber\\
&&U(|\Psi_2\rangle|P_0\rangle)=\sqrt {\gamma_2}|\Psi_2^\perp\rangle
\texttt{e}^{\texttt{i} \varphi}|P_0\rangle+\sqrt
{1-\gamma_2}|\Phi^{(2)}_{AP}\rangle.\nonumber\\
\end{eqnarray}
The linearity of $U$ implies that
\begin{eqnarray}\label{gaoyan6}
&&U(|\Psi_3\rangle|P_0\rangle)\nonumber\\ =&&(\alpha\sqrt
{\gamma_1}|\Psi_1^\perp\rangle+\beta\sqrt
{\gamma_2}|\Psi_2^\perp\rangle \texttt{e}^{\texttt{i}
\varphi})|P_0\rangle\nonumber\\
&&+\alpha\sqrt
{1-\gamma_1}|\Phi^{(1)}_{AP}\rangle+\beta\sqrt
{1-\gamma_2}|\Phi^{(2)}_{AP}\rangle.\end{eqnarray}
Hence, if
\begin{equation}\label{gaoyan66}
\alpha\sqrt {\gamma_1}|\Psi_1^\perp\rangle+\beta\sqrt
{\gamma_2}|\Psi_2^\perp\rangle \texttt{e}^{\texttt{i} \varphi}=\sqrt
{\gamma_3}|\Psi_3^\perp\rangle
\texttt{e}^{\texttt{i}\chi},\end{equation} one obtains
\begin{eqnarray}\label{gaoyan7}
\nonumber\\&& U(|\Psi_3\rangle|P_0\rangle)=\sqrt
{\gamma_3}|\Psi_3^\perp\rangle|P_0\rangle\texttt{e}^{\texttt{i}\chi}+\sqrt
{1-\gamma_3}|\Phi^{(3)}_{AP}\rangle.\nonumber\\
\end{eqnarray}
Here $\chi$ is a real number. Therefore,  we can realize the probabilistic perfect NOT transformation on the set $\{|\Psi_1\rangle, |\Psi_2\rangle, |\Psi_3\rangle\}$ in case of Eq.(\ref{gaoyan66}) being satisfied.

Now, suppose that
\begin{eqnarray}
|\Psi_1\rangle=\cos{\frac {\theta_1}{2}}|0\rangle+\sin{\frac
{\theta_1}{2}}\texttt{e}^{\texttt{i}\phi_1}|1\rangle,\nonumber\\
|\Psi_2\rangle=\cos{\frac {\theta_2}{2}}|0\rangle+\sin{\frac
{\theta_2}{2}}\texttt{e}^{\texttt{i}\phi_2}|1\rangle.
\end{eqnarray}
Eq.(\ref{gaoyan55}) then becomes
\begin{eqnarray}\label{gaoyan88}
&& |\alpha|^2+|\beta|^2+\alpha^*\beta(\cos{\frac
{\theta_1}{2}}\cos{\frac {\theta_2}{2}}+\sin{\frac
{\theta_1}{2}}\sin{\frac
{\theta_2}{2}}\texttt{e}^{\texttt{i}(\phi_2-\phi_1)})\nonumber\\
&&+\alpha\beta^*(\cos{\frac
{\theta_1}{2}}\cos{\frac {\theta_2}{2}}+\sin{\frac
{\theta_1}{2}}\sin{\frac
{\theta_2}{2}}\texttt{e}^{\texttt{-i}(\phi_2-\phi_1)})=1,\end{eqnarray}
 and Eq.(\ref{gaoyan66}) changes to
\begin{eqnarray}\label{gaoyan99}
&&\alpha\sqrt {\gamma_1}\cos{\frac
{\theta_1}{2}}+\beta\texttt{e}^{\texttt{i}\varphi}\sqrt
{\gamma_2}\cos{\frac {\theta_2}{2}}\nonumber\\
&&=\sqrt
{\gamma_3}\texttt{e}^{\texttt{i}\chi}(\alpha^*\cos{\frac
{\theta_1}{2}}+\beta^*\cos{\frac {\theta_2}{2}}), \nonumber\\
&&\alpha\sqrt {\gamma_1}\sin{\frac
{\theta_1}{2}}\texttt{e}^{\texttt{-i}\phi_1}+\beta\texttt{e}^{\texttt{i}\varphi}\sqrt
{\gamma_2}\sin{\frac
{\theta_2}{2}}\texttt{e}^{\texttt{-i}\phi_2}\nonumber\\
&&=\sqrt
{\gamma_3}\texttt{e}^{\texttt{i}\chi}(\alpha^*\sin{\frac
{\theta_1}{2}}\texttt{e}^{\texttt{-i}\phi_1}+\beta^*\sin{\frac
{\theta_2}{2}}\texttt{e}^{\texttt{-i}\phi_2}).\nonumber\\
\end{eqnarray}
Therefore, if Eqs.(\ref{gaoyan88}) and (\ref{gaoyan99}) are
satisfied by $\{|\Psi_1\rangle, |\Psi_2\rangle, |\Psi_3\rangle\}$,
then probabilistic perfect NOT transformation on this quantum state
set can be realized.

 For example, we can realize a probabilistic perfect NOT transformation with probability $\gamma$ on the  quantum states
 \begin{eqnarray}
 &&|\Psi_1\rangle=\frac {1}{\sqrt 2}(|0\rangle+|1\rangle),\nonumber\\
 &&|\Psi_2\rangle=\frac {1}{\sqrt 2}(|0\rangle+\texttt{i}|1\rangle),\nonumber\\
 &&|\Psi_3\rangle=\frac {1}{\sqrt 2}[(q+r \texttt{e}^{-\texttt{i}\frac {\varphi}{2}})|0\rangle
 +(q+\texttt{i} r \texttt{e}^{-\texttt{i}\frac {\varphi}{2}})|1\rangle)],\nonumber\\\end{eqnarray}
where $q,r, \gamma, \varphi$ are real and satisfy
\begin{equation}
\begin{array}{l}
q^2+r^2+\sqrt
2qr\cos(\frac {\pi}{4}-\frac {\varphi}{2})=1,\\
\frac {1}{2}-2\gamma+\frac {\gamma^2}{2}+\gamma\sin\varphi\geq 0,\\
1-\gamma\geq 0.\\
\end{array}
\end{equation}
Obviously,  $\varphi=\frac {\pi}{2}$ corresponds to the perfect NOT
transformation case.

\section{4. Conjugate transformation of a multi-level quantum system }

 In this section we discuss  conjugate transformation of a multi-level  quantum system (qudit). Suppose
 that the  dimension of a Hilbert space for  the quantum system is $d$. An arbitrary quantum state of the system
 can be  written as
 \begin{equation}
 |\Psi\rangle=\sum_{i=0}^{d-1}\alpha_i|i\rangle,\end{equation}
where $\alpha_i$ are complex numbers and $\{|i\rangle\}$ is an orthonormal basis. Let us define a conjugate
transformation $T$ as
\begin{equation}
T|\Psi\rangle=T(\sum_{i=0}^{d-1} \alpha_i|i\rangle)=\sum_{i=0}^{d-1}
\alpha^*_i|i\rangle\equiv |\Psi^T\rangle.
\end{equation}
Obviously, a perfect NOT transformation equals  $UT$  for a qubit,  where $
U=\left(\begin{array}{cc}0&-1\\
1&0\end{array}\right)$ is a unitary transformation. We call $|\Psi^T\rangle$ the conjugate state of quantum state
$|\Psi\rangle$. Evidently, one can not design a machine that will
take an arbitrary quantum state
 $|\Psi\rangle$ and transform it into its conjugate state $|\Psi^T\rangle$ because of the need
for complex conjugation of the coefficients in the transformation,
which must be accomplished by an antiunitary transformation and
cannot be performed by a unitary one. By Lemma, we can also assert
 that this kind transformation is impossible on a general quantum state set $S=\{|\Psi_1\rangle, |\Psi_2\rangle, \cdots, |\Psi_n\rangle\}$ of a qudit, since $\langle
\Psi_i|\Psi_j\rangle\neq \langle \Psi_i^T|\Psi_j^T\rangle=\langle
\Psi_i|\Psi_j\rangle^*$ for two arbitrary quantum states in the set $S$.

However, by  the  argument similar to the qubit case, we do have the following conclusions:

\emph{Theorem 1'.} A perfect conjugate
transformation
 on the state set $S=\{|\Psi_1\rangle, |\Psi_2\rangle,
\cdots, |\Psi_n\rangle\}$ of a qudit  can be realized by a unitary transformation if and only if all inner-products of the quantum
states in the set $S$ are real.

 \emph{Theorem 2'.}  Suppose that the quantum state set $S=\{|\Psi_1\rangle,
|\Psi_2\rangle, \cdots, |\Psi_n\rangle\}$ of a qudit satisfies
$\langle\Psi_i|\Psi_j\rangle\neq 0$. Then a conjugate transformation
 on the state set $S$ can be realized by a unitary transformation acting on the system and a probe if and only if Eq.(\ref{yangao1}) hold.

\emph{Theorem 3'}. The states $|\Psi_1\rangle, |\Psi_2\rangle, . . .,$ and $|\Psi_n\rangle$
  can be probabilistically perfectly transformed to their respective conjugate states if and only if there exist  a diagonal
  positive-definite matrix $\Gamma$ and $|P^{(i)}\rangle$ ($i=1,2,...,n$) such that   the matrix $X^{(1)}-\sqrt
 \Gamma X^{(T)}_P\sqrt {\Gamma^+}$ is positive-semidefinite. Here
$X^{(1)}=[\langle \Psi_i|\Psi_j\rangle]$ and
$X^{(T)}_P=[\langle \Psi_i^\perp|\Psi_j^\perp\rangle\langle
P^{(i)}|P^{(j)}\rangle]$ are $n\times n$ matrices, and  $|P^{(i)}\rangle$ $(i=1,2,\cdots,n)$ are quantum states of a probe.

\emph{Theorem 4'}. The states secretly chosen from the set $S=\{
|\Psi_1\rangle, |\Psi_2\rangle, \cdots, |\Psi_n\rangle\}$ of a qudit can be
probabilistically transformed into their respective conjugate states by a
general unitary-reduction operation if $
|\Psi_1\rangle, |\Psi_2\rangle, \cdots,$ and $|\Psi_n\rangle$ are
linearly independent.

Next we investigate the best possible efficiencies $\gamma_i$
attained by a probabilistic conjugate transformation.

For the sake of simplicity, we only discuss the special case   $S=\{|\Psi_1\rangle, |\Psi_2\rangle, |\Psi_3\rangle\}$, where $|\Psi_1\rangle, |\Psi_2\rangle, |\Psi_3\rangle$ are linearly independent. In this case, $X^{(1)}-\sqrt
 \Gamma X^{(T)}_P\sqrt {\Gamma^+}$ becomes
{\small\begin{widetext}
\begin{equation}\label{zhichao1}
\left(\begin{array}{ccc} 1-\gamma_1 &
\langle\Psi_1|\Psi_2\rangle-\sqrt {\gamma_1\gamma_2}\langle\Psi_1|\Psi_2\rangle^*\langle P^{(1)}|P^{(2)}\rangle & \langle\Psi_1|\Psi_3\rangle-\sqrt {\gamma_1 \gamma_3}\langle\Psi_1|\Psi_3\rangle^*\langle P^{(1)}|P^{(3)}\rangle\\
\langle\Psi_2|\Psi_1\rangle-\sqrt
{\gamma_1\gamma_2}\langle\Psi_2|\Psi_1\rangle^*\langle
P^{(2)}|P^{(1)}\rangle & 1- \gamma_2 &
\langle\Psi_2|\Psi_3\rangle-\sqrt {\gamma_2\gamma_3}\langle\Psi_2|\Psi_3\rangle^*\langle P^{(2)}|P^{(3)}\rangle\\
 \langle\Psi_3|\Psi_1\rangle-\sqrt
{\gamma_1\gamma_3}\langle\Psi_3|\Psi_1\rangle^*\langle
P^{(3)}|P^{(1)}\rangle & \langle\Psi_3|\Psi_2\rangle-\sqrt {
\gamma_2\gamma_3}\langle\Psi_3|\Psi_2\rangle^*\langle
P^{(3)}|P^{(2)}\rangle
 &  1-\gamma_3\end{array}\right).
\end{equation}
\end{widetext}}

Let  $\langle\Psi_1|\Psi_2\rangle=t_{12}\texttt{e}^{\texttt{i}\theta_{12}}$,
$\langle\Psi_1|\Psi_3\rangle=t_{13}\texttt{e}^{\texttt{i}\theta_{13}}$ and $\langle\Psi_2|\Psi_3\rangle=t_{23}\texttt{e}^{\texttt{i}\theta_{23}}$. We choose $\gamma_1=\gamma_2=\gamma_3=\gamma$,
$\langle P^{(1)}|P^{(2)}\rangle=\texttt{e}^{2\texttt{i}\theta_{12}}$ and $\langle P^{(1)}|P^{(3)}\rangle=\texttt{e}^{2\texttt{i}\theta_{13}}$. So Eq.(\ref{zhichao1}) becomes
\begin{equation}\label{zhichao6}
(1-\gamma)\left(\begin{array}{ccc} 1 &
\langle\Psi_1|\Psi_2\rangle & \langle\Psi_1|\Psi_3\rangle\\
 \langle\Psi_2|\Psi_1\rangle & 1 & A
\\
\langle\Psi_3|\Psi_1\rangle & A^*
 &  1\end{array}\right),
\end{equation}
where $A=\frac{\langle\Psi_2|\Psi_3\rangle-\gamma\langle\Psi_2|\Psi_3\rangle^*\texttt{e}^{2\texttt{i}(\theta_{13}-\theta_{12})}}{1-\gamma}$.
The positive-semidefinite condition of Eq.(\ref{zhichao6}) requires
\begin{equation}\label{gaoyan666}
\texttt{Det}\left(\begin{array}{ccc} 1 &
\langle\Psi_1|\Psi_2\rangle & \langle\Psi_1|\Psi_3\rangle\\
 \langle\Psi_2|\Psi_1\rangle & 1 & A
\\
\langle\Psi_3|\Psi_1\rangle & A^*
 &  1\end{array}\right)\geq 0, \end{equation}
 and
\begin{equation}\label{gaoyan888}
\texttt{Det}\left(\begin{array}{cc}
 1 & A
\\
 A^*
 &  1\end{array}\right)\geq 0. \end{equation}

Let $\delta=\theta_{12}-\theta_{13}+\theta_{23}$, $a=-\texttt{Det} X^{(1)}=-1+t_{12}^2+t_{13}^2+t_{23}^2-2t_{12}t_{13}t_{23}\cos\delta$, $b=t_{23}^2-1$.
 Eq.(\ref{gaoyan666}) means
 \begin{equation}
 a\gamma^2+2\gamma(2t_{23}^2\sin^2\delta-a)+a\leq 0.\end{equation}
So we have
 \begin{equation}
 0<\gamma\leq 1+\frac {2\sqrt{t_{23}^4\sin^4\delta-at_{23}^2\sin^2\delta  }-2t_{23}^2\sin^2\delta }{a}.
 \end{equation}

By Eq.(\ref{gaoyan888}) we obtain
\begin{equation}
 b\gamma^2+2\gamma(2t_{23}^2\sin^2\delta-b)+b\leq 0.\end{equation}
Therefore, $\gamma$  satisfies
\begin{equation}
 0<\gamma\leq 1+ \frac {2\sqrt{t_{23}^4\sin^4\delta-bt_{23}^2\sin^2\delta  }-2t_{23}^2\sin^2\delta}{b}.\end{equation}

 Since $a>b$, and  $1+2\frac{\sqrt {1-x}-1}{x}$ is a monotone function,  the maximum of $\gamma$ in this special case is
 \begin{equation}
 \gamma_{\text{max}}=1+\frac {2\sqrt{t_{23}^4\sin^4\delta-at_{23}^2\sin^2\delta  }-2t_{23}^2\sin^2\delta }{a}.
 \end{equation}

 Hence a lower bound of $\text{Max}(\frac {\gamma_1+\gamma_2+\gamma_3}{3})$ is $\gamma_{\text{max}}$.

\section{5. Summary }

In conclusion, we  have investigated a perfect NOT transformation on a quantum state set $S$ of a qubit and derived two necessary and sufficient conditions for realizing a perfect NOT transformation on $S$.
A probabilistic perfect NOT transformation (gate) was constructed by a general unitary-reduction operation. With a postselection of the measurement outcomes, the probabilistic NOT
gate was shown to yield perfect respective orthogonal complements of the input states.  We also show that
one can construct a probabilistic perfect NOT gate of the input
 states secretly chosen from a certain set $S=\{
|\Psi_1\rangle, |\Psi_2\rangle, \cdots, |\Psi_n\rangle\}$ if  $|\Psi_1\rangle, |\Psi_2\rangle, \cdots,$ and
$|\Psi_n\rangle$ are linearly independent. Furthermore, we  generalize the probabilistic NOT transformation to the conjugate transformation in a multi-level quantum system. The lower bound of the best possible efficiencies attained by a probabilistic perfect conjugate transformation was obtained.

\vspace{1cm}

\section{acknowledgments}
 We thank Professor M. D. Choi for helpful
discussions. This work was supported by the National Natural Science
Foundation of China under Grant No: 10971247, Hebei Natural Science
Foundation of China under Grant Nos: F2009000311, A2010000344.

\end{document}